\documentclass[sigconf]{acmart}

\AtBeginDocument{%
  }

\copyrightyear{2026}
\acmYear{2026}
\setcopyright{cc}
\setcctype{by}
\acmConference[ICSE '26]{2026 IEEE/ACM 48th International Conference on Software Engineering}{April 12--18, 2026}{Rio de Janeiro, Brazil}
\acmBooktitle{2026 IEEE/ACM 48th International Conference on Software Engineering (ICSE '26), April 12--18, 2026, Rio de Janeiro, Brazil}
\acmPrice{}
\acmDOI{10.1145/3744916.3787778}
\acmISBN{979-8-4007-2025-3/2026/04}

\usepackage{multirow}
\usepackage{rotating}
\usepackage[inkscapeformat=png]{svg}

\usepackage[utf8]{inputenc}
\usepackage[T1]{fontenc}    
\usepackage{url}

\usepackage{breakurl}

\usepackage{nicefrac}       
\usepackage{microtype}      
\usepackage{xcolor}         
\usepackage{multirow}
\usepackage{makecell}
\usepackage{subcaption}

\usepackage{xspace}
\usepackage{amsthm}
\usepackage{cancel}
\usepackage{color}
\usepackage{colortbl}

\usepackage{listings}
\usepackage{moreverb}
\usepackage{fancyvrb}

\usepackage{framed}
\usepackage{textcomp}
\usepackage{enumitem}  

\usepackage{soul}
\usepackage{hyperref}
\hypersetup{
    colorlinks=true,
    linkcolor=blue,
    citecolor=blue,
    filecolor=blue,      
    urlcolor=blue,
}

\usepackage{algorithm}
\usepackage{amsmath,amsfonts,bm}
\usepackage{setspace}
\usepackage{pifont}

\usepackage{xcolor}
\usepackage{framed}

\definecolor{emailgray}{gray}{0.95}
\newenvironment{EmailBoxBreakable}
{%
  \par\noindent
  \colorlet{shadecolor}{emailgray}%
  \begin{shaded}
  \raggedright

}
{%
  \end{shaded}
  \par
}

\begin{document}

\title{SpecOps: A Fully Automated AI Agent Testing Framework in Real-World GUI Environments}

\author{Syed Yusuf Ahmed}
\affiliation{%
  \institution{Purdue University}
  \city{West Lafayette}
  \state{Indiana}
  \country{USA}
}
\email{ahmed298@purdue.edu}

\author{Shiwei Feng}
\affiliation{%
  \institution{Purdue University}
  \city{West Lafayette}
  \state{Indiana}
  \country{USA}
}
\email{feng292@purdue.edu}

\author{Chanwoo Bae}
\affiliation{%
  \institution{Purdue University}
  \city{West Lafayette}
  \state{Indiana}
  \country{USA}
}
\email{bae68@purdue.edu}

\author{Calix Barrus}
\affiliation{%
  \institution{University of Texas at San Antonio}
  \city{San Antonio}
  \state{Texas}
  \country{USA}
}
\email{calix.barrus@my.utsa.edu}

\author{Xiangyu Zhang}
\affiliation{%
  \institution{Purdue University}
  \city{West Lafayette}
  \state{Indiana}
  \country{USA}
}
\email{xyzhang@cs.purdue.edu}

\renewcommand{\shortauthors}{Ahmed et al.}
\acmCodeLink{https://github.com/yusf1013/SpecOps}


\begin{abstract}
Autonomous AI agents powered by large language models (LLMs) are increasingly deployed in real-world applications, where reliable and robust behavior is critical. However, existing agent evaluation frameworks either rely heavily on manual efforts, operate within simulated environments, or lack focus on testing complex, multimodal, real-world agents. We introduce \textsc{SpecOps}\xspace, a novel, fully automated testing framework designed to evaluate GUI-based AI agents in real-world environments. \textsc{SpecOps}\xspace decomposes the testing process into four specialized phases—test case generation, environment setup, test execution, and validation—each handled by a distinct LLM-based specialist agent. This structured architecture addresses key challenges including end-to-end task coherence, robust error handling, and adaptability across diverse agent platforms including CLI tools, web apps, and browser extensions. In comprehensive evaluations across five diverse real-world agents, \textsc{SpecOps}\xspace outperforms baselines including general-purpose agentic systems such as AutoGPT and LLM-crafted automation scripts in planning accuracy, execution success, and bug detection effectiveness. \textsc{SpecOps}\xspace identifies 164 true bugs in the real-world agents with an F1 score of 0.89. With a cost of under \$0.73 and a runtime of under eight minutes per test, it demonstrates its practical viability and superiority in automated, real-world agent testing.
\end{abstract}


\maketitle

\vspace{-0.3cm}
\section{Introduction}

Autonomous Large Language Model (LLM) agents have rapidly evolved from simple text-generating systems to sophisticated real-world product-level agents capable of operating real-world applications. These agents---ranging from email assistants and HR chatbots to financial advisors and customer service representatives---can now navigate complex interfaces, execute multi-step tasks, operate on real data, and perform actions on behalf of users with minimal human intervention. As these systems transition from research prototypes to consumer products, they are becoming integral to critical business operations, handling sensitive financial transactions, managing customer relationships, and making decisions that directly impact organizational workflows and user experiences.

The widespread adoption of these agents in high-stakes environments makes robust testing methodologies essential to ensure their functionality, reliability, and safety. 
Product-level agents operate in dynamic, real-world environments where failures can have significant consequences---from sending incorrect emails to mishandling sensitive HR inquiries or corrupting important files. The rapid development cycles characteristic of AI systems, combined with frequently evolving requirements and system prompts, further amplify the need for comprehensive automated testing approaches.

Existing automated testing approaches have significant limitations that prevent effective evaluation of product-level agents. 
Current benchmarks require substantial manual effort to design test cases, create evaluation scripts, and interpret results. 
Many frameworks rely on simulated environments that fail to capture real-world interaction complexity and may suffer from simulator hallucinations that undermine test validity. 
Most importantly, existing approaches focus on toy agents with basic text-based input and simple tool access rather than sophisticated product-level agents requiring multimodal input, UI navigation, and advanced tool usage.

The fundamental limitation is that AI agents represent a new class of software that differs significantly from traditional systems. Unlike traditional software that operates on structured inputs with well-defined APIs and produces deterministic outputs, AI agents exhibit nondeterministic behaviors and operate with near free-form inputs and outputs. Testing such systems introduces unique challenges. In this work, we systematically outline these challenges and present solutions that have proven effective across several widely used agents. We integrate these solutions in \textsc{SpecOps}\xspace, a novel automated end-to-end testing framework for GUI-based AI agents.  Given the characteristics of AI agent systems being tested, which we call \textit{subject agents}, \textsc{SpecOps}\xspace employs a team of specialist LLM agents to automate the testing pipeline. Our approach adopts an {\em agent-like} design (i.e., an LLM orchestrating a suite of tools) for each specialist agent and makes extensive use of prompt engineering. Each specialist tackles distinct phases of the testing pipeline, yielding a complete end-to-end testing pipeline. 
Unlike previous approaches, \textsc{SpecOps}\xspace operates with minimal human input, autonomously generating comprehensive bug reports on subject agents under test that operate in real environments rather than simulators. \looseness=-1

In summary, we make the following contributions:
\begin{itemize}[leftmargin=1.5em]
    \item We propose \textsc{SpecOps}\xspace, the first fully automated, end-to-end testing framework for GUI-based, product-level LLM agents. 
    \item We design an architecture that (i) uses an adaptive strategy to react to constraints discovered during execution, (ii) employs specialist testing agents with tailored tools and guidelines to robustly interact with subject agents, and (iii) leverages human-like visual monitoring via screen captures to detect failures across diverse platforms (including CLI tools, web apps, and browser extensions).
    \item We implement and evaluate \textsc{SpecOps}\xspace on five diverse product-level agents spanning three domains (Email, File System, and HR Q\&A). \textsc{SpecOps}\xspace achieves a 100\% prompting success rate (compared to 11-49.5\% for baselines), perfect execution of planned steps, and identifies 164 true bugs with an F1 score of 0.89—significantly outperforming existing approaches while maintaining practical efficiency at under $\$0.73$ per test.
\end{itemize}

Our tools and findings from this study serve as a valuable resource for future research on automated agent testing. We share our code and data in the repository\footnote{\href{https://github.com/yusf1013/SpecOps}{https://github.com/yusf1013/SpecOps}}.

\begin{table*}[!th]
\centering
\small
\caption{Comparison of Related Work in Agent Testing (*Open source apps only)}
\label{tab:related_work}
\begin{center}
\vspace{-10pt}
\begin{tabular}{lcccc}
\toprule
\textbf{Related Work} & \textbf{Test Case Generation} & \textbf{Real-World Apps} & \textbf{Product-Level Agents} & \textbf{Validation} \\
\hline
OSWorld~\cite{xie2024osworld} & Manual & \checkmark & \texttimes & Manual Scripts   \\
AgentDojo~\cite{agentdojo} & Manual & \texttimes & \texttimes & Manual Scripts   \\
AgentCompany~\cite{agentcompany} & Manual & \checkmark* & \texttimes & Manual Scripts   \\
InjectAgent~\cite{injecagent} & Semi-automated & \texttimes & \texttimes & Manual  \\
R-Judge~\cite{rjudge} & Manual & \texttimes & \texttimes & Manual  \\
Windows Agent Arena~\cite{windowsagentarena} & Manual & \checkmark & \texttimes & Manual Scripts  \\
PrivacyLens~\cite{privacylens} & Semi-automated & \texttimes & \texttimes & Automated  \\
ToolEmu~\cite{toolemu} & Automated & \texttimes & \texttimes & Automated  \\
\hline
\textbf{\textsc{SpecOps}\xspace} & \textbf{Automated} & \textbf{\checkmark} & \textbf{\checkmark} & \textbf{Automated}  \\
\bottomrule
\end{tabular}%
\end{center}

\vspace{-8pt}
\end{table*}

\section{Related work}

\subsection{GUI Testing and Automation Tools}
Traditional GUI-testing research targets the reliability and usability of \emph{fixed} applications.
Random event generators (e.g., Android Monkey~\cite{monkeytest}) and model-based methods that build state-transition or event-flow graphs~\cite{nguyen2014guitar} systematically explore UI states, while vision-based or RL-driven tools predict widget hierarchies and learn input policies~\cite{seeing}.
Industrial suites such as Selenium~\cite{selenium}, Playwright~\cite{playwright}, and Cypress~\cite{cypress}, and newer vision-based~\cite{seeing} and LLM-based~\cite{kashef} products (Skyvern~\cite{skyvern}, TestSigma~\cite{testsigma}, TestRigor~\cite{testrigor}) automate scripted browser workflows.
LLM guided approaches (e.g., ScenGen~\cite{scengen}) have also automated GUI testing for mobile applications.
These approaches test traditional software (e.g., Android apps) with deterministic outputs for concrete inputs but do not generalize to LLM agents with non-deterministic behaviors arising from free-form natural language inputs and agentic decision-making. Additionally, they are limited within the target application (e.g., in the application GUI) and do not cover upstream tasks like constructing virtual user environments or downstream tasks like probing the environment for changes, which are critical for end-to-end testing.

\subsection{LLM Agent Safety Benchmarks}

Researchers have released a variety of suites that probe the robustness, alignment, and security of language-model agents in largely simulated environments.
ToolEmu~\cite{toolemu} and InjectAgent~\cite{injecagent} synthesize tasks or adversarial prompts on the fly, allowing large-scale robustness testing with minimal human effort.
AgentSafetyBench~\cite{agentsafetybench}, TrustAgent~\cite{trustagent}, AGrail~\cite{agrail}, AgentMonitor~\cite{agentmonitor}, and PrivacyLens~\cite{privacylens} measure whether agents follow safety policies or privacy regulatory constraints, scoring their responses to pre-defined scenarios.
Platforms such as AgentDojo~\cite{agentdojo} and R-Judge~\cite{rjudge} conduct multi-turn adversarial dialogues and assign risk labels to observed behaviors.
Although these benchmarks expose critical failure modes, they share two limitations: they run in text-only or simulated sandboxes, and they depend on human-curated labels or LLM-based simulators that may hallucinate operating-system interactions, thus restricting realism and scalability.

\subsection{Real-Environment Testbeds}
To close the realism gap, several frameworks embed agents in full operating systems.
OSWorld~\cite{osworld} pioneers cross-platform evaluation (Ubuntu, Windows, macOS) but requires $1\,{,}800+$ person-hours to curate $369$ tasks.
AgentCompany~\cite{agentcompany} simulates an entire software-company workflow with $3\,{,}000+$ design hours, and Windows Agent Arena~\cite{windowsagentarena} accelerates execution via Azure parallelization while still relying on handcrafted tasks.
These systems show that agents underperform in real settings, and the heavy cost of manual task curation limits the approaches' scalability.
In contrast, as shown in Table~\ref{tab:related_work}, \textsc{SpecOps}\xspace (i) boots agents in a \emph{real} operating environment, (ii) \emph{automatically} generates tasks and environment configurations, and (iii) validates functional correctness without human labels. To the best of our knowledge, it is the first framework that delivers end-to-end, fully automated testing for product-level, tool-using agents in realistic settings.\looseness=-1
\section{Motivation} 

In this section, we first present an example of testing a real-world agent (Section~\ref{sec:testing_real_agent}) and illustrate the limitations of existing methods (Section~\ref{sec:existing_limitation}). 
We then discuss the inherent challenges of the problem and introduce the design of our technique.

\subsection{Testing Real-World Agents} \label{sec:testing_real_agent}



We use a bug observed in Open Interpreter (OI)~\cite{oi}, a popular open-source agent (60k+ GitHub stars), as a motivating example to illustrate the limitations of baselines and highlight the need for a new approach.
OI runs on the user's local environment, so users can chat with it in natural language through the CLI. It is capable of general-purpose computing capabilities like creating and editing files (e.g., photos, videos, and PDFs).

When prompted to back up data from the current directory, OI mistakenly uses the absolute path (``\texttt{\textasciitilde/}'') instead of the correct relative path (``\texttt{./}''), causing a ``not found'' error.
Testing this functionality would involve creating target files to back up, navigating OI's interface to prompt it, monitoring its operation as it executes, probing the environment for changes, and detecting bugs. From an automation standpoint, this comes with numerous challenges:

\noindent \textbf{Challenge 1: End-to-end task coherence.} An end-to-end testing pipeline involves a chain of complex tasks. For instance, testing ``making backup" functionality involves setting up the environment by populating a target folder with data for backup, crafting a natural language prompt that mimics the style of a real-world user, controlling the test environment to activate and prompt the subject agent, monitoring its operation and output for completion indicators and errors, validating changes brought in the environment by the agent, and labeling bugs. Here, understanding long-term dependencies is crucial. For example, labeling bugs should be consistent with the environment setup and prompt designed in the earlier stages. It should allow any reasonable interpretation of the prompt in context, while checking implicit requirements like using proper language, formatting, and accurate communication with the user. All these require maintaining consistency throughout the entire pipeline.

\begin{figure*}[t]
    \centering
    \includegraphics[width=1.0\linewidth]{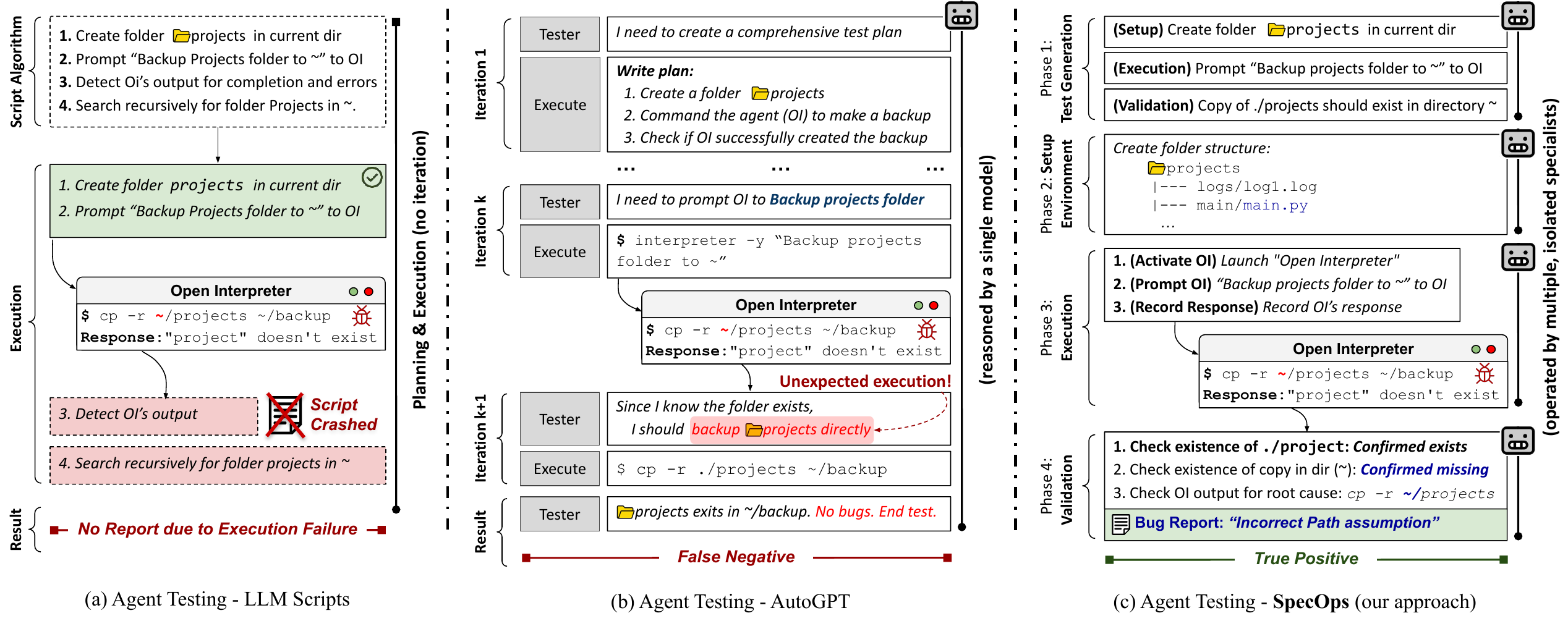}
    \caption{
    Different approaches to automate testing Open Interpreter's (OI) file backup functionality. 
    (a) LLM Scripts: As the test execution script encounters the buggy agent, it crashes. Lacking interactive design in testing, the system cannot fix the problem, leaving the remaining planned testing steps unattempted. 
    (b) AutoGPT: A single testing module carries out all test phases. Lack of testing isolation results in a failure to separate the subject agent's task (copying files) from AutoGPT's task (launching OI). 
    (c) SpecOps: Agent testing is divided into four distinct specialized phases. During the execution phase, \textsc{SpecOps}\xspace focuses solely on launching OI. In the validation phase, \textsc{SpecOps}\xspace detects the absence of backup files and successfully reports the bug. 
    } 
    \label{fig:baseline_failure}
    \vspace{-8pt}
    \Description{Motivating example}
\end{figure*}


\noindent \textbf{Challenge 2: Diversity.} 
Real-world subject agents operate across a wide spectrum of platforms, including web applications (e.g., ProxyAI \cite{proxyai}), browser extensions (e.g., TaxyAI \cite{taxyai}), command-line tools (e.g., Open Interpreter \cite{oi}), and hybrid desktop systems (e.g., Self-Operating Computer \cite{soc}). As a result, they have drastically different interfaces and modes of interaction.
Each setting imposes specific requirements on the testing agent for interaction, environment setup, and debugging.
Generic testing agents often lack the specialized logic or tools needed to handle these domain-specific differences, leading to fragile execution or complete test failures.

\noindent \textbf{Challenge 3: Lack of robustness and early failure detection.}
Agent testing must be efficient and fault-tolerant, especially when testing costly or time-consuming real-world agents. It needs to be robust enough to handle the diversity of subject agents. Inability to validate intermediate steps or recover from early mistakes can lead to cascading errors. Even small issues, like improper handling of an authentication session or flawed prompts, if undetected, can propagate through the entire execution, resulting in silent failures and wasted computation.

\subsection{Limitation of Existing Methods} \label{sec:existing_limitation}
Existing methods are not sufficient to effectively test real-world subject agents in an efficient and reliable manner.

\noindent \textbf{UI Testing Tools.} One might assume that traditional UI testing frameworks like Selenium~\cite{selenium} and Playwright~\cite{playwright} are sufficient to automate agent testing. While powerful for web automation, they are categorically unsuitable as baselines for agent testing because they require pre-scripted, deterministic workflows. These tools operate on fixed test scripts that must be manually crafted for specific UI elements and user flows, making them incapable of handling the dynamic, goal-oriented behavior of autonomous agents. More recent AI-powered testing tools like TestSigma~\cite{testsigma} and ScenGen~\cite{scengen} attempt to bridge this gap by incorporating LLM-driven automation, but they remain limited in scope—typically focusing exclusively on web or Android applications as opposed to CLI applications and browser extensions. Additionally, these tools specialize in workflow execution but fail to automate critical upstream and downstream phases such as test planning and environment setup and probing. This leaves substantial manual effort in designing test scenarios, configuring realistic testing environments, and interpreting results.

\noindent \textbf{LLM Script.} A natural next step is to leverage an LLM to automate that manual effort. One could prompt an LLM with a complete testing task, asking it to generate a static script that interacts with the subject agent. While easy to implement, this approach lacks mechanisms for runtime monitoring, requiring all test logic to be scripted upfront. This demands robust upfront knowledge of how the agent will behave, which is unrealistic given agents' inherent non-determinism. Even minor deviations from anticipated behavior such as an unexpected response format or slightly different action sequence can cause the script to crash entirely, invalidating the entire test. \looseness=-1


Figure~\ref{fig:baseline_failure} illustrates attempts to automate the testing of OI's file backup operation functionality. For ease of presentation, we assume the same planned setup across all approaches—creating a backup for the folder ``projects''. Figure~\ref{fig:baseline_failure}~(a) shows the inadequacy of LLM scripts when applied to real-world testing. The script encounters a bug from the subject agent and crashes as a result. Consequently, the rest of the planned testing cannot proceed, leaving the bug undetected. This demonstrates how Challenge~2 can be relevant even for CLI applications, where subtle operational flows can crash the whole testing framework, leaving downstream tasks incomplete.

\noindent \textbf{General-Purpose Agentic System.}
A natural way to overcome this lack of adaptability is to incorporate a feedback loop, essentially using an LLM agent for testing.
Agentic systems (e.g., AutoGPT~\cite{yang2023auto}) are more sophisticated, employing iterative reasoning, tool usage, and self-reflection. These systems are better equipped to react to dynamic feedback compared to static scripts.
However, they are not designed or optimized for testing, especially in real-world domains where test efficiency, fault isolation, and minimal execution waste are vital.
A minor early-stage error, if not promptly detected and corrected, can render the entire test invalid, wasting both time and resources. Without explicit support for test assertions, checkpoints, or failure recovery, these systems remain fragile and cost-inefficient for real-world agent testing.

As shown in Figure \ref{fig:baseline_failure} (b), AutoGPT starts with a correct plan in iteration 1, demonstrating a good understanding of the task. Initially, this understanding is reflected in the implementation, as it prompts OI correctly. However, upon receiving an error response from OI in iteration 3, it confuses its task, derailing from the plan. It misinterprets the objective as creating the backup itself rather than testing OI's capability to do so. Instead of identifying and reporting OI's behavior as buggy, it treats the response as an error message and attempts a direct method of creating a backup copy. This demonstrates AutoGPT's failure to address Challenge 1, where it loses consistency of steps along the pipeline, unable to recover from errors.\looseness=-1

\subsection{Our Approach} \label{sec:overview_challenges}

To address these challenges, we propose \textsc{SpecOps}\xspace, a structured, specialist-agent testing framework. As shown in Figure \ref{fig:baseline_failure}~(c),
\textsc{SpecOps}\xspace decomposes the testing process into four phases: {\em Test Generation}, {\em Environment Setup}, {\em Execution}, and {\em Validation}. 
Each phase is handled by a \emph{specialist}: an autonomous agent whose tools, heuristics, and objectives are tuned exclusively to that phase.

A naive multi-phase or multi-agent approach to agent testing would still fail, as early-stage errors (e.g., in environment setup) often only manifest in later phases, causing cascading failures (Challenge~1). To address this, we propose \emph{adaptive strategy}, where we employ a dual-specialist approach to produce a bundled test specification that ensures coherence across environment setup, test prompts, and validation oracles. This specification is updated during execution to reflect newly discovered constraints (e.g., environment limitations). Each subsequent phase references this evolving specification, maintaining consistency throughout the pipeline.

Translating this specification to execution requires interacting with the diverse interfaces of real-world agents (Challenge~2). We address this using UI interaction as a universal abstraction for any on-screen agent interface—employing keyboard/mouse primitives for control and screen recording analysis for monitoring. Additionally, the execution workflow must be fault-tolerant (Challenge~3). Unlike LLM Scripts (which often crash), \textsc{SpecOps}\xspace resolves any encountered errors within their respective phases via feedback loops, enabling retries or adaptation and minimizing error propagation. 

In the motivating example, Figure~\ref{fig:baseline_failure} shows that the goal of the execution phase is limited to delivering the designed prompt to OI. The resulting observation (i.e., no copied files) is propagated to the Validation phase, which is operated by a different specialist. By allowing each specialist to focus on a specific task and setting tight boundaries on its action space, we prevent the hallucination issue exhibited in AutoGPT (i.e., fixing the bug itself) and instead successfully report the bug.
\section{\textsc{SpecOps}\xspace Design}

\begin{figure*}[t]
\centering
\includegraphics[width=\textwidth]{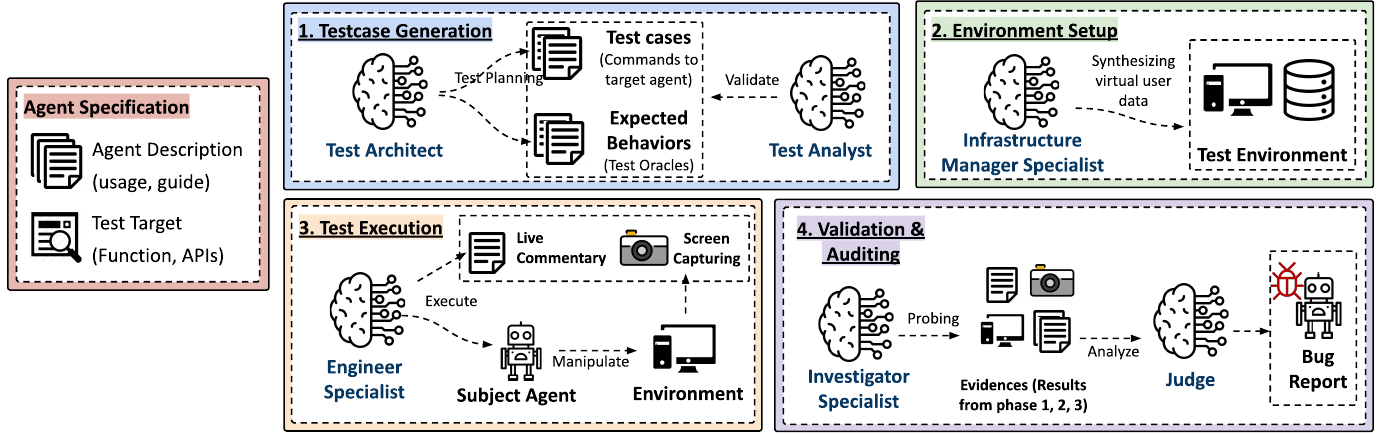}
\caption{\textsc{SpecOps}\xspace architecture showing the four-phase testing workflow with specialized agents operating each phase and Agent Specification as the input.}
\label{fig:main_figure}
\vspace{-11pt}
\Description{Architecture of SpecOps}
\end{figure*}

\subsection{Overview and Architecture}

The core idea behind \textsc{SpecOps}\xspace is to decompose complex, multi-step agent testing into a series of small, focused tasks. Each testing phase is specialized, equipped with domain-specific expertise and tools tailored to its particular role.

Figure~\ref{fig:main_figure} presents the workflow of \textsc{SpecOps}\xspace. In this section, we introduce the details of each testing phase. To illustrate how \textsc{SpecOps}\xspace conducts agent testing, we present a running example: testing the ``email reply'' feature in ProxyAI. ProxyAI is a commercial multi-agent system that operates through the collaboration of dynamically created worker agents.

\subsection{Phase 1: Test Case Generation}
The test case generation phase transforms high-level feature descriptions into detailed, executable test scenarios. Given an abstract feature, this phase designs the environment setup, the prompt for the subject agent, and the expected behavior by breaking the feature down into specific testable scenarios that reflect real-world usage. Unlike conventional software with objective validation criteria, agent testing operates entirely in natural language—features, prompts, and expected behaviors are all linguistically specified. This presents three key challenges.

\noindent \textbf{Natural Language Variability.} With input prompts in natural language, infinite variations are possible. A good test case must identify relevant implicit expectations (``don't use placeholders, be professional, include specific dates...'') without over-specifying, while including appropriate checks for those expectations in the agent's output.

\noindent \textbf{Prompt Completeness.} As identified in PrivacyLens~\cite{privacylens}, prompts must not lack crucial details that prevent the subject agent from executing the task, which would move testing out of scope. 

\noindent \textbf{Oracle Generalizability.} Measurable outcomes (oracles) must allow for all valid execution paths. For example, to ``find the email from David,'' an agent might use the search bar or scroll through visible emails. Oracles checking only one approach will flag valid alternatives as false positives.

We address these challenges through \textit{two-specialist} \textit{adaptive strategy} based on modified self-reflection. Research shows that asking LLMs to reflect on their own answers helps identify errors~\cite{shinn2023reflexion,pan2024automatically,madaan2023selfrefine,toy2024metacognition,wang2024metacognitive,asai2024selfrag}. Here, we find better results by separating the guidelines between the generation and reflection phases. The generation phase focuses on prompt completeness and naturalness, while reflection re-evaluates these along with feasibility of environment setup, and completeness and generalizability of oracle checks. We name the former the Test Architect agent and the latter the Test Analyst agent. \looseness=-1

\textbf{Running Example}: Given the email reply feature, the Test Architect generates: 
\begin{itemize}
    \item Environment Setup: One email needs to be added to the inbox from David asking for the quarterly projections
    \item Prompt: Reply to David's email from yesterday with ``Q3 projections''.
    \item Expected behavior: A reply to David's email on 2025-05-12 should be sent. The content in the reply should contain ``Q3 projections'' information. 
\end{itemize}

The Test Analyst examines the initial test case and identifies a critical deficiency: the prompt instructs the subject agent to reply with Q3 projections, but that data does not exist. Hence, it adds an extra step to the environment setup to create ``Q3 projection data''. $\Box$ \looseness=-1

\subsection{Phase 2: Environment Setup}
Agent testing requires the construction of virtual user environments, which serve as an additional layer of the input space. \textsc{SpecOps}\xspace addresses this by introducing a dedicated phase for generating user-level testing environments populated with virtual user data.

The key challenge in this phase is testing a diverse range of features with minimal API access to real-world environments. Unlike tests using simulated environments or self-hosted open-source apps, API access requirements for similar data manipulation capabilities are much higher with real-world systems like Gmail. For file system testing, we operate as a non-root user, restricted to the \texttt{/home/user} directory. For email testing, we have access to only a single API function that sends a fresh email from a fixed domain (@aibrilliance.online) to the target inbox, with no API access to the target inbox itself. This limits environment manipulation capabilities (e.g., changing sender email domains or timestamps). \looseness=-1

To address these limitations, the Infrastructure Manager specialist breaks down high-level environment setup requirements from Phase 1 into concrete tool invocations using MCP~\cite{mcp}, observes tool call outputs, and applies chain-of-thought reasoning to make necessary adjustments to the test case's prompt and expected behavior. \looseness=-1

This design makes our approach platform-agnostic, eliminating dependencies on specific service providers. For example, we do not depend on Gmail's proprietary APIs—our email testing works with any email provider that supports basic message delivery. We demonstrate in our evaluation that this minimal API setup can indeed support testing diverse features from existing benchmarks.

\textbf{Running Example:} The Infrastructure Manager decides the specific content of David's email and uses the API call to populate the Gmail account.

\begin{EmailBoxBreakable}
Email from: David Peterson\\
Date: 2025-05-12, Time: 17:10\\
Subject: Project Timeline Questions\\
Email body:\\[1em]
Dear John,\\
Could you please provide an update on the Q3 projections?\\[1em]
Thanks.
\end{EmailBoxBreakable}
In the generated email, the name, date, and content are aligned with the testing context. It also generates plausible Q3 projection data. $\Box$ \looseness=-1

\subsection{Phase 3: Test Execution}
In this phase, agent testing is conducted by the Engineer Specialist in the \textsc{SpecOps}\xspace framework. It first gathers the test cases (i.e., commands directed at the subject agent) and the corresponding testing environment, then proceeds to execute the agent. 

The first key challenge to address is interacting with agents across diverse platforms. To achieve generalizability across the diverse UIs of different agents, we develop UI-based tools that abstract platform-specific differences into universal interaction primitives: keyboard and mouse controls. We implement specialized versions of these as MCP tools for the Engineer specialist. Our enhanced keyboard tool verifies with certainty if the typed text actually appears on the screen. This tells the Engineer to retry when a key press does not correspond to new text appearing on the screen (e.g., when typing is attempted without selecting an input field). For mouse interaction, we adapt approaches similar to SOM~\cite{som} but focus on text-based clicking rather than highlighting. When identical text appears in multiple screen locations, we use relative positioning parameters that help the agent disambiguate targets.

The next key challenge is comprehensive output monitoring. Capturing and preserving the full runtime activity of the subject agent is crucial, as it serves as reliable evidence for detecting bugs and diagnosing their root causes. In most product-level GUI agents, key runtime traces can only be captured visually (i.e., via the screen). Hence, \textsc{SpecOps}\xspace introduces a novel design choice, leveraging screen captures to record the agent’s execution. Whenever the screen changes, \textsc{SpecOps}\xspace captures the runtime display, which is then forwarded as key evidence for potential bugs. 

\textbf{Running Example}: The Engineer uses the browser to navigate to ProxyAI (the subject agent) and launches it with the prompt (\texttt{``Reply to David’s email from yesterday with Q3 project-\\ions, which are...''}). It then monitors the agent's execution and logs/captures its observations. $\Box$

\subsection{Phase 4: Validation and Auditing}
At this stage, \textsc{SpecOps}\xspace aggregates all results from the previous phases to investigate and detect bugs. The main challenge lies in consolidating diverse types of evidence (e.g., text, images, and environment data) which combined, creates a heavy burden of information processing, resulting in hallucinations. To address this, we employ \textit{Human-like Visual Monitoring}. First, we separate the task into focused specialists: the Judge, who reasons over the collected information to identify bugs, and the Investigator, who supports the Judge by gathering data from multiple sources, probing the environment to discover any status changes and presenting findings in organized text. This reduces the inherent reasoning complexity for the Judge. Second, we use the Meta Chain of Thoughts (Meta-CoT) prompting method~\cite{mikepradel,zhou2022least}, which we discuss below.

\noindent\textbf{Investigator Specialist.} The Investigator is deployed when agent actions modify the testing environment, requiring independent verification of environmental changes.
Similar to the Infrastructure Manager (Phase 2), the Investigator interacts directly with the test environment; however, its primary focus is on detecting and reporting status changes resulting from the subject agent's execution. \looseness = -1

\noindent\textbf{Judge} The Judge examines all available evidence: commentary, screen captures, and status changes in environment and compares them with test oracles to identify bugs. To accurately identify bugs, we instruct it to report when one of the following conditions is met: \looseness=-1
\begin{itemize}\label{text:bugdef}
    \item Unreasonable deviation from expected behavior  
    \item Misreporting by the agent to users
    \item Impact on completion of intended outcome
    \item Impact on quality of intended outcome  
    \item Requiring unreasonable user intervention
\end{itemize}
The Judge needs to address various sources of information, ranging from images to text. To reduce hallucination, we employ Meta-CoT prompting strategy. Inspired by prior work~\cite{cot,zhou2022least}, our Meta-CoT prompting first asks the Judge to generate CoT questions on the available information and answer the CoT questions itself. Example prompts are available on our online repository.

\textbf{Running Example}: The Investigator checks the Gmail Inbox (i.e., sent folder) and navigates to the thread containing David's email. It then discovers the reply email containing the requested Q3 projections.  
The Judge analyzes the evidence and finds that the agent signed off the email with a placeholder value ``[your name]'' without replacing it with the corresponding name (i.e., David) $\Box$

\section{Evaluation} 
To evaluate the effectiveness of \textsc{SpecOps}\xspace, we performed experiments spanning five diverse GUI-based product-level agents (whose features cover three application domains) and two baseline LLM-powered test automation approaches.
In each application domain, we systematically selected specific features that should be supported by the corresponding agents (e.g., an HR chatbot should be able to answer questions about employee sick leave policy). 

\subsection{Subject Agents}
To show the generalizability of \textsc{SpecOps}\xspace, we selected subject agents with different architectures, interaction paradigms, and application domains.
The subject agents and their respective application domains are presented in Table \ref{tab:agent_listing}.
The five subject agents are:
\begin{itemize}[leftmargin=1.5em]
    \item \textbf{ProxyAI~\cite{proxyai}:} A commercial web-based agent (\$20/month subscription) with a closed-source architecture that operates through dynamic worker agents, recently acquired by Salesforce~\cite{salesforce2025acquisition}.
    \item \textbf{Open Interpreter (OI)~\cite{oi}:} An open-source agent with $59{,}000+$ GitHub stars and $3{,}000+$ commits. 
    \item \textbf{Self-Operating Computer (SOC)~\cite{soc}:} An open-source agent from HyperWriteAI with $600+$ commits and $10{,}000+$ GitHub stars.
    \item \textbf{TaxyAI~\cite{taxyai}:} An open-source browser plugin with $1.2$k stars on GitHub.
    \item \textbf{Autonomous HR Chatbot (HRC)~\cite{ahc}:} An open-source chatbot with around $400$ stars and $62$ commits.
\end{itemize}

\subsubsection{Subject Agent Application Domains}
The selected subject agents have testable features from three domains:
\begin{itemize}[leftmargin=1.5em]
    \item \textbf{HR:} HRC has testable capabilities related to queries about Human Resources data. These features usually involve: 1) CRUD operations on an employee database, and 2) Q\&A from a given HR Policy document.
    \item \textbf{Email:} ProxyAI, SOC, and TaxyAI have testable capabilities involving email-related tasks (e.g., ``find specific email'').
    \item \textbf{File System:} OI has testable capabilities on file system operations (e.g., ``find files by name'').
\end{itemize}

\subsubsection{Tested Feature Selection} 
For each domain, we construct a feature list that should be supported by the corresponding subject agents.
We systematically construct this list through a multi-stage process combining automated feature extraction and manual curation: \looseness=-1

\noindent \textbf{Stage 1 - Automated Feature Extraction:} 
We employ an LLM to analyze official documentation (GitHub repositories, websites, API docs) for each subject agent to identify claimed capabilities and features. This yields the initial feature list. \looseness=-1

\noindent \textbf{Stage 2 - Cross-benchmark feature mining:} 
We create a separate list of features by extracting the underlying testable (and relevant) features from test cases in the following agent-evaluation benchmarks: OSWorld~\cite{osworld}, AGrail~\cite{agrail}, AgentSafetyBench (ASB)~\cite{agentsafetybench}, ToolEmu~\cite{toolemu}, and WorkBench~\cite{workbench}. 
We reduce this list of relevant features by removing redundant features and features not supported by our selected subject agents. 
The number of selected benchmark features grouped by benchmark is presented in Table \ref{tab:feature_extraction} and aggregated in Table \ref{tab:domain_features} (\# Extracted Benchmark Features). 

\begin{table}[!t]
\centering
\small
\caption{Feature extraction across benchmarks}\label{tab:feature_extraction}
 \vspace{-10pt}
\begin{tabular}{lccc}
\toprule
\textbf{Benchmark} & \textbf{\makecell{\# Test \\ Cases}} & \textbf{\makecell{\# Relevant \\Features}} & \textbf{\makecell{\# Selected \\Features}} \\
\hline
\makecell{OSWorld \\(File system)} & 24 & 10 & 8 \\
\hline
ASB (email) & 114 & 17 & 13 \\
\hline
ASB (file) & 31 & 17 & 16 \\
\hline
\makecell{ToolEmu \\(email)} & \multirow{2}{*}{51} & 9 & 8 \\
\cline{1-1} \cline{3-4}
\makecell{ToolEmu \\(file)} & & 15 & 12 \\
\hline
\makecell{AGrail \\(file)} & 47 & 24 & 22 \\
\hline
\makecell{WorkBench \\(email)} & 92 & 9 & 8 \\
\bottomrule
\end{tabular}
\vspace{-13pt}
\end{table}

\noindent \textbf{Stage 3 - Manual Augmentation:} 
We add a subset of the selected benchmark features to our initial list where the feature:
\begin{itemize}[leftmargin=1.5em]
    \item is not sufficiently covered by already included features
    \item can be feasibly tested considering implementation complexity and environmental constraints
\end{itemize}
In Table \ref{tab:domain_features}, we show the number of features in our final selection (\# Features in Final Selection) as well as the number of extracted benchmark features covered by this final selection. To be considered covered, a benchmark feature must map to at least one in the final list. \looseness=-1

For the HR domain, existing benchmarks lack Q\&A-focused test cases. We therefore supplement the feature list using MongoDB and SQL reference materials for database CRUD operations, along with HR policy documentation.

\begin{table}[!t]
    \centering
    \caption{Statistics of feature extraction}
    \label{tab:domain_features}
     \vspace{-10pt}
    \begin{tabular}{cccc}
        \toprule
        \textbf{Domain} & \textbf{\makecell{\# Extracted \\ Benchmark \\ Features}} & \textbf{\makecell{\# Benchmark \\ Features \\ Covered}} & \textbf{\makecell{\# Features \\ in Final \\ Selection}} \\
        \hline
        Email & 25 & 20 & 17 \\
        File System & 44 & 39 & 20 \\
        \bottomrule
    \end{tabular}
\end{table}

\begin{table}[!t]
    \centering
    \small
    \caption{Subject agents, corresponding application domains, and count of testable features examined by \textsc{SpecOps}\xspace and baselines}
    \label{tab:agent_listing}
     \vspace{-10pt}
    \begin{tabular}{ccc}
        \toprule
        \textbf{Domain} & \makecell{\textbf{Subject Agents}} & \makecell{\textbf{\# Features}} \\
        \hline
        Email & ProxyAI, SOC, TaxyAI & 17 \\
        File System & OI & 20 \\
        HR Q\&A & AHC & 29 \\
        \bottomrule
    \end{tabular}
    \vspace{-8pt}
\end{table}

\subsection{Baselines}
To the best of our knowledge, no existing framework directly addresses the same problem as \textsc{SpecOps}\xspace (i.e., testing real-world AI agents). We follow the comparison approach from the most relevant work~\cite{mikepradel} and pick two representative baselines that embody alternative automation strategies and are the most applicable to our problem setting: LLM Scripts and AutoGPT. 
The details of these baselines can be found in Section~\ref{sec:existing_limitation}.

\noindent \textbf{Baseline 1: LLM Scripts.} Following~\cite{mikepradel}, we implement this baseline by asking an LLM to generate a script (e.g., Python, Bash) that performs end-to-end testing on a specific feature.
The LLM receives comprehensive context including: 
\begin{enumerate}[leftmargin=1.5em]
    \item The specific feature to be tested
    \item Detailed agent documentation and execution instructions
    \item Screenshots demonstrating a complete sample run of the agent
    \item API specifications and usage instructions for environment setup
    \item A standardized starting environment configuration
\end{enumerate}
Generated scripts utilize Playwright~\cite{playwright} for browser automation, providing access to standard web testing functionality including element interaction, page navigation, and content verification. 
The scripts are responsible for all four testing phases: environment setup, agent activation and prompting, oracle verification, and cleanup.

\noindent \textbf{Baseline 2: AutoGPT.} 
Following~\cite{mikepradel}, our second baseline evaluates whether general-purpose autonomous agents can effectively handle testing tasks without specialized design. 
AutoGPT is a powerful general-purpose tool with access to specialized tools like web search, file I/O operations, Python code execution, terminal access, etc. 
It provides iterative planning and execution capabilities, allowing it to refine its approach based on intermediate results.
Following the paper's configuration, we allow AutoGPT a maximum of 50 iterations with a 10-minute timeout per iteration step (a maximum of 500 minute budget for each test case). 
However, since AutoGPT cannot process image inputs, we replace screenshots with detailed textual descriptions generated by an LLM. 
We manually verify that these descriptions capture all essential visual information present in the original screenshots, ensuring no information loss that would unfairly disadvantage this baseline.

\vspace{-8pt}
\subsection{Research Questions} 

For our evaluation experiments, we use each testing framework to look for bugs in all subject agents. 
Each testing framework generates and executes one test for each testable feature.

To compare \textsc{SpecOps}\xspace with the baseline frameworks, we propose the following research questions:

\noindent \textbf{RQ1 (Test Planning):} How do the planning capabilities of \textsc{SpecOps}\xspace compare to baselines? 

To answer RQ1, we first discretize the testing process proposed and generated by each framework into a listing of steps.
Details on this discretization are discussed in Section~\ref{sec:discretization}.

Using this discretization, human annotators then evaluate: 

\textbf{Incorrect Steps:} 
Planned steps that are technically infeasible, procedurally wrong, or would not achieve their intended purpose. 
Incorrect steps indicate fundamental misunderstandings of the testing environment or agent behavior (e.g., clicking on an incorrect location when attempting to access an input field).
For this metric, steps are counted as incorrect if they are incorrect \emph{as written}; i.e., the steps are identifiable as incorrect to human annotators before execution. \looseness=-1

\textbf{Missing Steps:} The number of necessary steps not included in the plan. 
These counts of incorrect and missing steps indicate how often a testing framework makes fundamental planning errors—decisions that are easily spotted as wrong upon cursory manual inspection.

\noindent \textbf{RQ2 (Execution Ability):} How successfully does \textsc{SpecOps}\xspace execute planned tests compared to the baselines? 

To answer RQ2, we start by using the same step discretization as RQ1.
We then measure how effectively each approach implements its planned steps in practice by reporting the number of planned steps that are correctly executed in the actual testing environment.
We collect this data by using traces and artifacts from the executed test and having human annotators count how many of the steps succeeded.
Steps dependent on failed predecessors are also counted as execution failures.
These counts of successful test execution steps indicate how often testing frameworks fail to execute their proposed plan due to technical failures.

\noindent \textbf{RQ3 (Effectiveness):} How effectively does \textsc{SpecOps}\xspace find and identify bugs in subject agents compared to baselines?

To answer RQ3, we collect all bugs in subject agents reported by testing frameworks.
We then manually classify these bugs as: 

\textbf{True Positives (TP):} Bugs in the subject agent correctly identified by both the evaluator and the human annotators on the same execution trace.

\textbf{False Positives (FP):} Behaviors reported as bugs by the evaluator but classified as benign by human annotators.
Note that if a bug is reported but is caused by an environment setup failure on the part of the testing framework (and not caused by the subject agent), it is labeled as an FP.

We also review the test executions to identify \textbf{False Negatives (FN):} bugs observed in the subject agent's execution and identified as such by human annotators, but missed by the evaluator.

In the event that technical shortcomings prevent the test framework from successfully communicating with the subject agent (i.e., for the given test, the framework is unable to get a prompt to the subject agent), any reported bugs are considered FPs and reported separately below.

\noindent \textbf{RQ4 (Failure Cause Analysis):} What are the root causes of failures in automated agent testing frameworks?

Having established performance differences across planning, execution, and bug detection (RQ1-3), we conduct a root cause analysis to understand the underlying factors that contribute to these differences. This analysis provides insights into common failure modes and the impact of our architectural design choices.

To answer RQ4, we examine all observed failures across the three frameworks, which manifest as incorrect or missing test steps (RQ1), execution failures (RQ2), and false positives/negatives (RQ3). We manually analyze execution logs, error traces, and generated artifacts for all test cases to identify the underlying causes of these failures. \looseness=-1

\noindent \textbf{RQ5 (Efficiency):} What is the time/token efficiency of \textsc{SpecOps}\xspace in performing agent testing? 

To answer RQ5, we monitor the time and tokens consumed by \textsc{SpecOps}\xspace during the experiments.
We use this information to compute the estimated cost of our approach. 

\subsubsection{Step Discretization}
\label{sec:discretization}

During initial experimentation, we found that treating tests or test phases as units did not allow us to isolate sources of failure or provide granular insight into each approach.
To this end, we systematically decompose the tests produced by each framework into steps.
This discretization allows us to give ``partial credit'' when evaluating planning quality and execution capability, allowing for more insightful analysis.

To begin, we decompose the testing process into four phases: \textbf{environment setup, subject agent activation and prompting, oracle verification,} and \textbf{environment cleanup}. 
In our presentation, we omit cleanup since it is not applicable for all frameworks or application domains.

\noindent \textbf{Step Counting Criteria:} 
To ensure consistency across evaluators using different automation frameworks, we decompose generated tests into distinct, discrete steps using the following criteria:

\textit{UI tasks}
\vspace{-3pt}
\begin{itemize}
    \item \textbf{Navigation operations:} Each UI state transition required to access verification targets (e.g., inbox $\rightarrow$ sent folder $\rightarrow$ specific email thread).
    \item \textbf{Data extraction:} Each conceptually distinct piece of information retrieved for comparison (e.g., email recipient, subject line, and attachment count are separate extractions).
\end{itemize}

\textit{Non-UI tasks}
\vspace{-3pt}
    \begin{itemize}
        \item Email creation: 4 steps without attachment + 1 step for each attachment.
        \item File creation with specific content: 2 steps (create, populate).
        \item File/folder creation with no content (touch, mkdir): 1 step.
        \item Terminal commands chained with \&, |, or similar operators: counted separately.
        \item All other terminal commands: 1 step each.
    \end{itemize}
    
\textit{Natural language:} Planning statements or descriptions of intended actions without further breakdown or execution attempts are counted as presented.

The breakdown of steps is given in Table~\ref{tab:discretized}.

\begin{table}[]
\centering
\small
\caption{Total count of discretized steps in Environment Setup, Test Execution, and Validation phases of framework-generated tests.}
\label{tab:discretized}
\vspace{-10pt}
{\renewcommand{\arraystretch}{0.85}
\begin{tabular}{llcccc|c}
\toprule
 & \textbf{Agent} & \textbf{\# Tests} & \textbf{Setup} & \textbf{Execution} & \textbf{Validation} & \textbf{Sum} \\
\hline
\multirow{6}{*}{\rotatebox{90}{LLM Scripts}} 
& Proxy & 17 & 91 & 110 & 133 & 334 \\
& SOC & 17 & 79 & 147 & 157 & 383  \\
& TaxyAI & 16 & 58 & 204 & 150 & 412 \\
& OI & 20 & 197 & 132 & 243 & 572 \\
& HRC & 29 & - & 255 & 325 & 580 \\
\cmidrule{2-7}
& \textbf{Total} & 99 & 425 & 848 & 1,008 & 2,281  \\
\hline

\multirow{6}{*}{\rotatebox{90}{AutoGPT}} 
& Proxy & 17 & 55 & 105 & 115 & 275  \\
& SOC & 17 & 61 & 94 & 88 & 243 \\
& TaxyAI & 16 & 38 & 116 & 85 & 239 \\
& OI & 20 & 41 & 85 & 100 & 226 \\
& HRC &  29 &  -    &  134    &    108 & 242 \\
\cmidrule{2-7}
& \textbf{Total} & 99 & 195 & 534 & 496 & 1,125 \\
\hline

\multirow{6}{*}{\rotatebox{90}{\textsc{SpecOps}\xspace}} 
& Proxy & 17 & 113 & 85 & 379 & 577 \\
& SOC & 17 & 119 & 102 & 248 & 469 \\
& TaxyAI & 16 & 115 & 144 & 307 & 566 \\
& OI & 20 & 163 & 100 & 300 & 563 \\
& HRC & 29 & - & 87 & 381 & 468 \\
\cmidrule{2-7}
& \textbf{Total} & 99 & 510 & 518 & 1,615 & 2,643 \\
\bottomrule
\end{tabular}
}
\vspace{-14pt}
\end{table}

\subsection{Environment Configuration}

All experiments are conducted on standardized hardware configurations to ensure comparable results. 
Each testing environment consists of an Ubuntu platform with a Chrome browser pre-configured with the debug port enabled. 
We assume all necessary authentication credentials are saved in the browser. 
This setup eliminates authentication overhead while providing realistic application access.
We use Claude 3.7 Sonnet (the latest version available during this research) as our base LLM model. We pair it with generalizable prompt templates for each specialist agent, filling only the input for each of the 99 test cases. 

\noindent \textbf{Application Setup:} 
For email domain testing, we use Gmail accounts with saved login credentials. 
File system testing operates on a standard Ubuntu file system. 
The HR chatbot testing environment includes access to the chatbot's web interface running on a local server. 
We use the default HR policy documents as a base environment but expand the default employee database to 25 employees.

\noindent \textbf{API Access:} We provide minimal environment modification capabilities through custom APIs: 
\begin{enumerate}
    \item $send\_email$ API that adds (sends) a regular email to the test Gmail account
    \item Command-line access for file system operations (no sudo)
\end{enumerate}

\subsection{RQ1: Test Planning Result}

\begin{table}[]
\centering
\small
\caption{Count of framework-generated test steps that are Incorrect as planned (\# I) or Missing (\# M). Percentages (\% I) are out of total count of discretized steps (Table~\ref{tab:discretized}).}
\label{tab:planning}
\vspace{-10pt}
\setlength{\tabcolsep}{4pt} 
{\renewcommand{\arraystretch}{0.9}
\begin{tabular}{llccccccccc}
\toprule
& \textbf{Agent} &
\multicolumn{3}{c}{\textbf{Setup}} &
\multicolumn{3}{c}{\textbf{Execution}} &
\multicolumn{3}{c}{\textbf{Validation}} \\
\cmidrule(lr){3-5} \cmidrule(lr){6-8} \cmidrule(lr){9-11}
& & \# I & \% I & \# M & \# I & \% I & \# M & \# I & \% I & \# M \\
\hline
\multirow{5}{*}{\rotatebox{90}{LLM Scripts}} 
& Proxy  & 0 & 0.0\% & 0 & 1 & 0.9\% & 1 & 0 & 0.0\% & 20 \\
& SOC    & 0 & 0.0\% & 0 & 5 & 3.4\% & 0 & 0 & 0.0\% & 25 \\
& TaxyAI & 5 & 8.6\% & 2 & 9 & 4.4\% & 0 & 1 & 0.7\% & 45 \\
& OI     & 1 & 0.5\% & 0 & 0 & 0.0\% & 0 & 0 & 0.0\% & 0 \\
& HRC    & - & - & - & 0 & 0.0\% & 0 & 0 & 0.0\% & 0 \\
\cline{2-11}
& \textbf{Total}    & 6 & 1.4\% & 2 & 15 & 1.8\% & 1 & 1 & 0.1\% & 90 \\
\hline
\multirow{5}{*}{\rotatebox{90}{AutoGPT}} 
& Proxy  & 0 & 0.0\% & 14 & 2 & 1.9\% & 1 & 6 & 5.2\% & 9 \\
& SOC    & 1 & 1.6\% & 6 & 8 & 8.5\% & 21 & 1 & 1.1\% & 16 \\
& TaxyAI & 0 & 0.0\% & 8 & 3 & 2.6\% & 2 & 1 & 1.2\% & 2 \\
& OI     & 3 & 7.3\% & 17 & 1 & 1.2\% & 10 & 0 & 0.0\% & 6 \\
& HRC    & - & - & - & 2 & 1.5\% & 0 & 5 & 4.8\% & 4 \\
\cline{2-11}
& \textbf{Total}    & 4 & 2.1\% & 45 & 16 & 3.0\% & 34 & 13 & 2.6\% & 37 \\
\hline
\multirow{5}{*}{\rotatebox{90}{\textsc{SpecOps}\xspace}} 
& Proxy  & 0 & 0.0\% & 0 & 0 & 0.0\% & 0 & 0 & 0.0\% & 1 \\
& SOC    & 0 & 0.0\% & 0 & 0 & 0.0\% & 0 & 0 & 0.0\% & 1 \\
& TaxyAI & 0 & 0.0\% & 0 & 0 & 0.0\% & 0 & 0 & 0.0\% & 22 \\
& OI     & 0 & 0.0\% & 0 & 0 & 0.0\% & 0 & 0 & 0.0\% & 0 \\
& HRC    & - & - & - & 0 & 0.0\% & 0 & 0 & 0.0\% & 0 \\
\cline{2-11}
& \textbf{Total}    & 0 & 0.0\% & 0 & 0 & 0.0\% & 0 & 0 & 0.0\% & 24 \\
\bottomrule
\end{tabular}
}
\vspace{-14pt}
\end{table}

The results for RQ1 are presented in Table~\ref{tab:planning}.
\textsc{SpecOps}\xspace outperforms both baselines across all phases, with no incorrect steps during the setup and execution phases, and significantly fewer missing steps overall.
LLM Scripts achieve near-perfect planning for most agents in setup, having 0 incorrect and missing steps for Proxy and SOC, and only 1 incorrect step for OI. It maintains a similar trend in planning the execution phase, achieving near-perfect numbers for three agents (OI, AHC, Proxy) and demonstrating a good understanding of CLI and web interfaces. However, it struggles significantly with handling TaxyAI's browser extension interface. In planning validation steps, LLM Scripts perform poorly, missing a total of 90 steps. The root cause is frequently omitted environment verification steps, relying instead on agent self-reports. This is a critical oversight given that agents often misreport their actions.
AutoGPT, despite being a more capable agent, suffers from a higher number of incorrect and missing steps in setup and execution. In the validation phase, it misses fewer steps than LLM Scripts in total (37 vs. 90) but is more prone to choosing incorrect steps (13 vs. 1).  

\textsc{SpecOps}\xspace outperforms both baselines, demonstrating the effectiveness of dual-specialist approach. It eliminates planning errors entirely from setup and execution, and has near perfect scores for 4 agents in validation. However, it misses 22 steps with Taxy. Manual inspection reveals that these have no downstream effect as in these cases, the Judge specialist already had sufficient information from previous phases to draw correct conclusions.

\subsection{RQ2: Success Rate on Test Execution}

\begin{table}[]
\small
\centering
\setlength{\tabcolsep}{3pt} 
\caption{Test Execution success rates with raw counts (successes/attempts and percentiles) by phase}
\label{tab:execution}
\vspace{-10pt}
\begin{tabular}{llccc}
\toprule
& \textbf{Agent} & \textbf{Setup} & \textbf{Execution} & \textbf{Validation} \\
\hline
\multirow{6}{*}{\rotatebox{90}{LLM Scripts}} 
& Proxy  & 100.0\% (91/91)   & 42.7\% (47/110)  & 16.5\% (22/133) \\
& SOC    & 100.0\% (79/79)   & 83.0\% (122/147) & 41.4\% (65/157) \\
& TaxyAI & 100.0\% (58/58)   & 18.1\% (37/204)  & 1.3\% (2/150) \\
& OI     & 94.4\% (186/197)  & 73.5\% (97/132)  & 55.6\% (135/243) \\
& HRC    & --                & 60.4\% (154/255) & 8.0\% (26/325) \\
\cline{2-5}
& \textbf{Total}  & \textbf{97.6\% (414/425)}  & \textbf{53.9\% (457/848)} & \textbf{24.8\% (250/1008)} \\
\hline
\multirow{6}{*}{\rotatebox{90}{AutoGPT}} 
& Proxy  & 92.7\% (51/55)    & 22.9\% (24/105)  & 0.0\% (0/115) \\
& SOC    & 100.0\% (61/61)   & 46.8\% (44/94)   & 0.0\% (0/88) \\
& TaxyAI & 89.5\% (34/38)    & 18.1\% (21/116)  & 1.2\% (1/85) \\
& OI     & 82.9\% (34/41)    & 51.8\% (44/85)   & 24.0\% (24/100) \\
& HRC    & --                & 22.4\% (30/134)  & 0.0\% (0/108) \\
\cline{2-5}
& \textbf{Total}  & \textbf{92.3\% (180/195)}  & \textbf{33.5\% (163/534)} & \textbf{5.0\% (25/496)} \\
\hline
\multirow{6}{*}{\rotatebox{90}{SpecOps}} 
& Proxy  & 100.0\% (113/113) & 100.0\% (85/85)   & 96.3\% (365/379) \\
& SOC    & 100.0\% (119/119) & 100.0\% (102/102) & 89.5\% (222/248) \\
& TaxyAI & 100.0\% (115/115) & 100.0\% (144/144) & 94.5\% (290/307) \\
& OI     & 100.0\% (163/163) & 100.0\% (100/100) & 99.7\% (299/300) \\
& HRC    & --                & 100.0\% (87/87)   & 98.4\% (375/381) \\
\cline{2-5}
& \textbf{Total}  & \textbf{100.0\% (510/510)} & \textbf{100.0\% (518/518)} & \textbf{96.0\% (1551/1615)} \\
\bottomrule
\end{tabular}
\end{table}

The results for RQ2 are presented in Table~\ref{tab:execution}. All three approaches show good execution capability in setting up the environment, which completely comprises of making API calls and executing terminal commands. However, in the next two phases, \textsc{SpecOps}\xspace far outperforms the baselines. It is able to implement 100\% and 96\% of its planned Execution and Validation steps, respectively. LLM Scripts could only implement 53.9\% and 24.8\%, while AutoGPT remained at 33.5\% and 5\%.
For both baselines, we frequently observe failures to communicate with the subject agent, which is one of the main causes of poor execution outcomes, causing cascading failures. \looseness=-1

\subsection{RQ3: Effectiveness on Bug Detection}

\begin{table}[]
    \centering
    \begin{minipage}{\linewidth}
        \caption{Evaluation Results: Bug Detection Performance}
        \vspace{-10pt}
        \label{tab:bug_det}
        
        \Description{A detailed table showing bug detection results for five agents: Proxy, SOC, Taxy, OI, and HRC. It compares three methods: LLM Scripts, AutoGPT, and SpecOps. For each agent and method, it lists the number of tests, True Positives (TP), False Positives (FP), and False Negatives (FN). SpecOps consistently shows higher True Positive counts across most agents compared to LLM Scripts and AutoGPT.}
        
        \small
        \centering
        \setlength{\tabcolsep}{2pt} 
        
        \begin{tabular}{cccccccccccccc}
            \toprule
            \multirow{2}{*}{\textbf{Agent}} & \multirow{2}{*}{\textbf{Case}} & \multicolumn{4}{c}{\textbf{LLM Scripts}} & \multicolumn{4}{c}{\textbf{AutoGPT}} & \multicolumn{4}{c}{\textbf{\textsc{SpecOps}\xspace}} \\
            \cmidrule(lr){3-6} \cmidrule(lr){7-10} \cmidrule(lr){11-14}
            & & Tests & TP & FP & FN & Tests & TP & FP & FN & Tests & TP & FP & FN \\
            \midrule 
            
            \multirow{3}{*}{Proxy} & SP & 9 & 1 & 16 & 3 & 0 & - & - & - & 17 & 28 & 3 & 5 \\
            & UP & 8 & - & 5 & - & 17 & - & - & - & 0 & - & - & - \\
            \cline{2-14}
            & \textbf{Sum} & \textbf{17} & \textbf{1} & \textbf{21} & \textbf{3} & \textbf{17} & \textbf{-} & \textbf{-} & \textbf{-} & \textbf{17} & \textbf{28} & \textbf{3} & \textbf{5} \\
            \hline
            
            \multirow{3}{*}{SOC} & SP & 12 & 11 & 5 & 27 & 2 & 0 & 0 & 8 & 17 & 47 & 4 & 14 \\
            & UP & 5 & - & 7 & - & 15 & - & 0 & - & 0 & - & - & - \\
            \cline{2-14}
            & \textbf{Total} & \textbf{17} & \textbf{11} & \textbf{12} & \textbf{27} & \textbf{17} & \textbf{0} & \textbf{0} & \textbf{8} & \textbf{17} & \textbf{47} & \textbf{4} & \textbf{14} \\
            \hline
            
            \multirow{3}{*}{Taxy} & SP & 0 & - & - & - & 1 & 0 & 0 & 1 & 16 & 53 & 1 & 7 \\
            & UP & 16 & - & 14 & - & 15 & - & 0 & - & 0 & - & - & - \\
            \cline{2-14}
            & \textbf{Sum} & \textbf{16} & \textbf{0} & \textbf{14} & \textbf{0} & \textbf{16} & \textbf{0} & \textbf{0} & \textbf{1} & \textbf{16} & \textbf{53} & \textbf{1} & \textbf{7} \\
            \hline

            \multirow{3}{*}{OI} & SP & 11 & 1 & 7 & 1 & 7 & 0 & 0 & 1 & 20 & 13 & 1 & 0 \\
            & UP & 9 & - & 3 & - & 13 & - & 0 & - & 0 & - & - & - \\
            \cline{2-14}
            & \textbf{Sum} & \textbf{20} & \textbf{1} & \textbf{10} & \textbf{1} & \textbf{20} & \textbf{0} & \textbf{0} & \textbf{1} & \textbf{20} & \textbf{13} & \textbf{1} & \textbf{0} \\
            \hline

            \multirow{3}{*}{HRC} & SP & 17 & 0 & 1 & 0 & 1 & 0 & 0 & 1 & 29 & 23 & 6 & 0 \\
            & UP & 12 & - & - & - & 28 & - & 0 & - & 0 & - & - & - \\
            \cline{2-14}
            & \textbf{Sum} & \textbf{29} & \textbf{0} & \textbf{1} & \textbf{0} & \textbf{29} & \textbf{0} & \textbf{0} & \textbf{1} & \textbf{29} & \textbf{23} & \textbf{6} & \textbf{0} \\
            \hline
            \hline
            \multirow{1}{*}{\textbf{Total}} & \textbf{-}  & \textbf{99} & \textbf{13} & \textbf{58} & \textbf{31} & \textbf{99} & \textbf{0} & \textbf{0} & \textbf{11} & \textbf{99} & \textbf{164} & \textbf{15} & \textbf{26} \\
            \bottomrule
        \end{tabular}
        
        \footnotesize 
        \vspace{2pt}
        \textit SP: Successful Prompt. \quad \quad UP: Unsuccessful Prompt
    \end{minipage}
\end{table}



\begin{table}[]
    \centering
    \begin{minipage}{0.8\columnwidth} 
        \caption{Aggregate Bug Detection Statistics} \label{tab:agg}
        \vspace{-10pt}
        \small
        \centering 
        {\renewcommand{\arraystretch}{0.85}
        \begin{tabular}{cccc}
            \toprule
             \textbf{}  & \textbf{LLM Script} & \textbf{AutoGPT} & \textbf{\textsc{SpecOps}\xspace}  \\
             \midrule
             PSR  & 49.5 \% & 11.1\% &  100\%  \\
             Bugs triggered & 44 & 11 & 190 \\
             Precision  & 0.18 & - & 0.92  \\
             Recall  & 0.30 & 0 & 0.86  \\
             F1 Score  & 0.23 & - & 0.89  \\
             \bottomrule
        \end{tabular}
        }
        \vspace{2pt}
        \footnotesize 
        PSR: Prompting Success Rate. \quad Bugs triggered: $TP + FN$
    \end{minipage}
\end{table}

Table~\ref{tab:bug_det} shows the bug detection performance of each approach across different agents, with Table~\ref{tab:agg} summarizing the aggregate results. \looseness=-1
\textsc{SpecOps}\xspace drastically outperforms the baselines. The Prompting Success Rate (PSR) indicates how frequently an evaluator was able to submit the intended prompt to the subject agent. Both baselines struggle here, with LLM Scripts only achieving 49.5\% and AutoGPT achieving 11.1\%, as opposed to a 100\% PSR for \textsc{SpecOps}\xspace. 
\textsc{SpecOps}\xspace was also able to trigger and detect the majority of true positive bugs in the subject agents, totaling 164, as opposed to 13 true positives by LLM Scripts. 
AutoGPT was unable to label any bugs, yielding no true positives or false positives. Hence, its precision and F1 score are undefined. \textsc{SpecOps}\xspace demonstrates impressive performance by achieving an F1 score of 0.89, while LLM Scripts only achieved 0.23. \looseness=-1

\subsection{RQ4: Failure Cause Analysis}
\begin{figure}[t]
    \centering
    \includegraphics[width=0.8\linewidth]{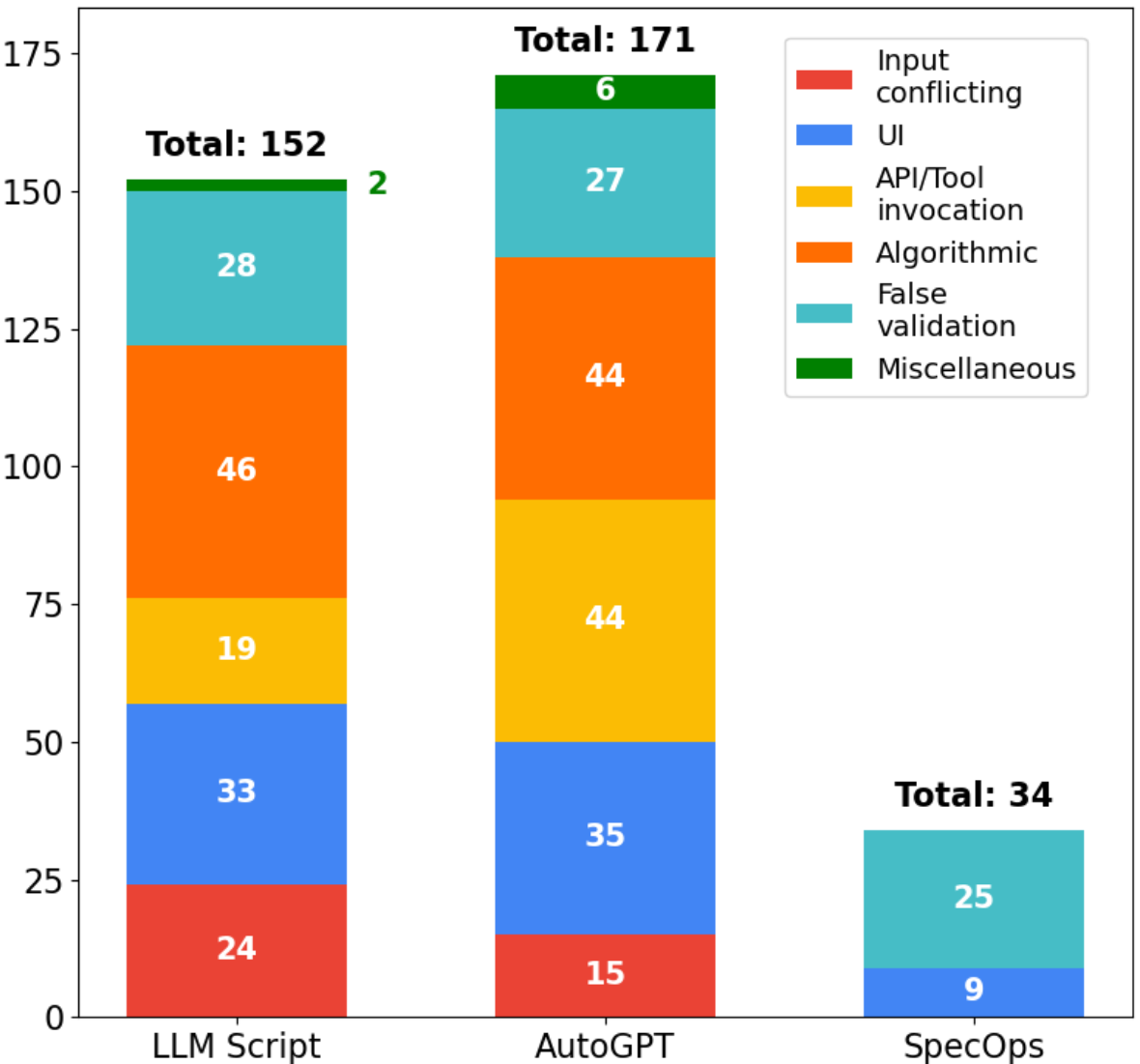}
    \caption{Root causes of failure across all systems}
    \label{fig:rca}
    \vspace{-10pt}
    \Description{Graph}
\end{figure}


From our analysis, we find that all observed failures of testing systems stem from hallucinations, which we categorize into 6 types:
\textcircled{1} Input-conflicting: Actions contradicting the input objective (e.g., fixing the subject agent's bugs rather than reporting them).\\
\textcircled{2}  UI: Hallucinating non-existent UI elements or identifiers.\\
\textcircled{3}  API/Tool invocation: Calling non-existent or irrelevant APIs/tools or with non-existent parameters.\\
\textcircled{4}  Algorithmic: Imagining steps or implementing them in a way that seems plausible but fails in execution. E.g., using \texttt{os.system('cd /tmp')} then \texttt{os.system('ls')} expecting to list \texttt{/tmp}, but each call runs in separate shell sessions.\\
\textcircled{5}  False validation: Assertion checks that seem plausible but prove to be brittle, insufficient, or unjustified in execution.\\
\textcircled{6}  Miscellaneous: Other hallucinations.

\smallskip

Figure~\ref{fig:rca} shows hallucination counts by type. \textsc{SpecOps}\xspace exhibits drastically fewer hallucinations (34 total) compared to LLM Scripts (152) and AutoGPT (171), and successfully avoids input-conflicting, API/tool, and algorithmic hallucinations.

\noindent \textbf{Baselines.} Baseline failures often stem from inability to distinguish testing from task execution and their over reliance on narrow assumptions rather than accounting for valid behavioral variations. These manifest across hallucination types: 
\textcircled{2} UI hallucinations lack robust handling for alternate UI presentations and 
\textcircled{4} algorithmic hallucinations implement brittle interaction logic that crashes on any deviation. 
Critically, \textcircled{5} false validation hallucinations demonstrate a lack of oracle generalizability, accepting very specific subject agent outcomes but breaking for valid alternatives.
Ideally, AutoGPT's iterative execution should allow recovery from both types of failures.
However, in practice, iteration without testing domain knowledge compounds the issue. 
Rather than diagnosing root causes, it attempts irrelevant fixes such as increasing wait timeouts for non-existent UI elements instead of inspecting the UI. 
In many cases, it invokes irrelevant command-line tools discovered through web searches that it lacks the capabilities to interact with, creating permanent deadlocks (\textcircled{3} tool invocation hallucinations). 
Moreover, \textcircled{1} input-conflicting hallucinations reveal conceptual confusion: LLM Scripts report their own scripting errors as bugs in the subject agent, while AutoGPT treats subject agents' error messages as problems to solve rather than behaviors to report. 
Both baselines collapse the distinction between tester and testee, treating them as components of a task-completion pipeline rather than separate entities in a testing relationship.

\noindent \textbf{Our Mitigation.} \textsc{SpecOps}\xspace exhibits no observed failures from 
\textcircled{1} input-conflicting, \textcircled{3} API/tool, or \textcircled{4} algorithmic hallucinations, and mitigates failures from others entirely through architectural design principles that make extensive use of feedback and validation loops to encode testing expertise. 
For example, the Test Analyst validates the Test Architect's specification before execution to catch \textcircled{1} input conflicting errors. 
Each specialist follows the specification to guide reasoning, minimizing \textcircled{4} algorithmic hallucinations arising from contextual inconsistencies common in long-horizon, multi-stage reasoning. 
Designated tools for each specialist come with built-in verification, detecting errors (e.g., failed typing) and providing actionable feedback (e.g., ``select input field'') to guide recovery from \textcircled{2} UI and \textcircled{3} tool invocation hallucinations. 
These mechanisms help all 99 cases reach validation, unlike baselines which frequently crashed or were blocked in earlier stages. 
Yet, \textsc{SpecOps}\xspace controls \textcircled{5} validation hallucinations through human-like visual monitoring. The Judge's Meta-CoT prompting generated multiple questions per bug type, systematically surfacing relevant details and guiding structured reasoning. \looseness=-1

\vspace{-5pt}
\subsection{RQ5: Efficiency of Agent Testing}

\begin{table}[t]
    \centering
    \small
    \captionof{table}{\textsc{SpecOps}\xspace mean times, tokens, and costs across tests} \label{tab:cost}
    \vspace{-10pt}
    {\renewcommand{\arraystretch}{0.8}
    \begin{tabular}{ccccccc}
        \toprule
         \textbf{Agent}  & \makecell{\textbf{End to End}\\\textbf{Runtime (s)}} & \makecell{\textbf{Input}\\\textbf{Tokens}} & \makecell{\textbf{Output}\\\textbf{Tokens}} & \textbf{Cost (\$)} \\
         \midrule
         Proxy  & 686 & 289.6K &  12.5K & 1.06 \\
         Taxy  & 496 & 203.8K & 11.5K & 0.77 \\
         SOC  & 742 & 445.1K & 160.0K & 1.58 \\
         OI  & 299 & 97.9K & 8.4K & 0.42 \\
         HRC & 221 & 32.8K & 7.0K & 0.20 \\
         \bottomrule
    \end{tabular}
    }
    \vspace{-13pt}
\end{table}

The results for RQ5 are presented in Table~\ref{tab:cost}.
During our experiments, our approach cost roughly one USD or less for 4 out of 5 subject agents. The average across all 99 tests approaches \$0.73 per test, resulting in a total budget of around \$72. The end-to-end runtime totals 12.5 hours for all 99 tests, which is less than 8 minutes per test. \looseness=-1

\vspace{-5pt}
\subsection{Discussion}

\noindent \textbf{Test diversity}
A key aspect of evaluating end-to-end testing approaches is tackling the inherent diversity in generated test cases. 
Owing to their architectural differences and inherent non-determinism, different testing systems make different test case design and implementation choices, which could impact downstream tasks such as the number of bugs triggered. 
The impact of these differences can be seen in the different numbers of planned steps reported in Table~\ref{tab:discretized}.
In these step counts, step discretization with the same objective metrics yields vastly different numbers of distinct planned steps for each tester (AutoGPT: 1,225, LLM Script: 2,281, and \textsc{SpecOps}\xspace: 2,643 steps total).
This difference persists across phases (e.g., 496 vs. 1,008 vs. 1,615 steps respectively in validation) and across subject agents (e.g., 242, 580, and 468 total steps respectively for HRC). 
These planning and implementation differences correlate with vastly different bug detection outcomes as shown in Tables~\ref{tab:bug_det} and \ref{tab:agg} (e.g., 11, 44, and 190 total bugs triggered respectively).

To ensure fairness and consistency in comparison despite these variations, we provide identical inputs to all testing systems (same feature descriptions, same agent documentation, same API access) and compare them using the same objective criteria and metrics (RQ1-RQ4), allowing us to assess each approach's effectiveness on its own terms while maintaining experimental validity.

\noindent \textbf{Environment Setup Failures} 
Since comprehensive agent testing requires setting up certain preconditions in environment setup, failure during that setup can invalidate the test.
This can result in false positives (e.g. a behavior is reported as a bug, but a precondition wasn't met) or false negatives (e.g. the failure to delete a file is misinterpreted, since the file was never there in the first place).
Environment setup failures were observed in both baselines, especially AutoGPT, but never observed in SpecOps during our evaluation (Table~\ref{tab:planning},\ref{tab:execution}, Col. ``Setup'').
We attribute SpecOps' avoidance of such failures to its dedicated environment setup phase.
Utilizing tool call output in feedback loop, this phase enables \textsc{SpecOps}\xspace to either retry \& recover, or clearly distinguish environment setup failures from subject agent failures \& abort gracefully. 
As a sanity check, we ran \textsc{SpecOps}\xspace, simulating a total of 20 environment setup failures by manually injecting such cases: internet disconnection (Gmail), shortage of storage (file system), and API timeouts.
In these tests, the Infrastructure Manager read the error in API call responses, retried, and after the set number of retries, abandoned execution without reporting bugs, citing environment failure.
This check shows that \textsc{SpecOps}\xspace' environment setup phase makes it possible for it to exit gracefully in the face of environment setup failure.

\vspace{-5pt}
\section{Conclusion}
This paper presents \textsc{SpecOps}\xspace, a fully automated testing framework for real-world GUI-based LLM agents that leverages a specialist-agent architecture to decompose testing into distinct phases. \textsc{SpecOps}\xspace achieves a 100\% prompting success rate, perfect execution of planned steps, and identifies 164 true bugs with an F1 score of 0.89, significantly outperforming baselines while maintaining practical efficiency at \$0.73 per test. This work establishes a foundation for the scalable, automated testing of increasingly sophisticated AI agents deployed in production environments.

\begin{acks}
 We are grateful to the Center for AI Safety for providing computational resources. This work was
funded in part by the National Science Foundation (NSF) Awards SHF-1901242, SHF-1910300,
Proto-OKN 2333736, IIS-2416835, DARPA VSPELLS-HR001120S0058, ONR N00014-23-1-2081,
and Amazon. Any opinions, findings and conclusions or recommendations expressed in this material
are those of the authors and do not necessarily reflect the views of the sponsors
\end{acks}

\clearpage
\bibliographystyle{ACM-Reference-Format}
\bibliography{reference}

\end{document}